\documentclass[12pt]{article}
\usepackage{amssymb,amsmath}
\usepackage{cite}
\newlength{\dinwidth}
\newlength{\dinmargin}
\newtheorem{theorem}{Theorem}[section]

\newtheorem{definition}[theorem]{Definition}

\newcommand{\ie}{{\it i.e.\ }}
\newcommand{\eg}{{\it e.g.\ }}
\newcommand{\etc}{{\it etc}}

\def\idty{{\leavevmode\hbox{\rm 1\kern -.3em I}}}

\def\As{{\cal A}}
\def\Bs{{\cal B}}

\def\Fs{{\cal F}}

\def\Hs{{\cal H}}

\def\Ks{{\cal K}}

\def\Ms{{\cal M}}
\def\Ns{{\cal N}}
\def\Os{{\cal O}}
\def\Ps{{\cal P}}

\def\Zs{{\cal Z}}

\def\Pid{{\Ps_+ ^{\uparrow}}}

\def\pair{{(\As,\Bs)}}
\def\pairf{{(\As(\Os_1),\As(\Os_2))}}
\def\idty{{\leavevmode\hbox{\rm 1\kern -.3em I}}}

\def\RR{{\mathbb R}}
\def\CC{{\mathbb C}}

\def\IN{{\mathbb N}}

\def\C{$C^{\ast}$-}
\def\W{$W^{\ast}$-}
\def\tr{{\rm Tr}}

\def\beq{\begin{equation}}
\def\eeq{\end{equation}}
\begin{document}
\title{Subsystems and Independence in Relativistic Microscopic Physics
\thanks{This is an expanded version of an invited talk
given at the biennial meeting of the Philosophy of Science Association, held 
in Pittsburgh, PA, on November 6--9, 2008. }
}
\author{{\Large Stephen J.\ Summers\,}\\[5mm]
Department of Mathematics \\
University of Florida \\ Gainesville FL 32611, USA}

\date{February 15, 2009}

\maketitle 

{\abstract \noindent The analyzability of the universe into
subsystems requires a concept of the ``independence'' of the
subsystems, of which the relativistic quantum world supports many
distinct notions which either coincide or are trivial in the classical 
setting. The multitude of such notions and the complex relations between 
them will only be adumbrated here. The emphasis of the discussion is 
placed upon the warrant for and the consequences of a particular notion 
of subsystem independence, which, it is proposed, should be viewed as 
primary and, it is argued, provides a reasonable framework within which to 
sensibly speak of relativistic quantum subsystems.}

\section{Introduction} \label{intro}

     Without the possibility of analyzing the universe into
subsystems, it is hardly conceivable how the sciences could be carried
out. Common experience certainly supports this possibility; however,
common experience is neither quantum nor relativistic, so it is far
from obvious whether one can sensibly speak of microscopic subsystems,
despite the fact that much of science is carried out as if one could
do so. Indeed, what appears to be straightforward in a classical,
nonrelativistic setting turns out to be highly nontrivial and, to some
degree, impossible in a relativistic quantum setting. However, it is
not our purpose here to rehearse the well known controversies
concerning the various kinds of nonlocality and interdependence of
subsystems manifest in relativistic quantum theory
(cf. \cite{But,ClHa,Suvac,PeTe} for recent discussions and
reviews). Instead, our intent is to indicate how, in spite of these,
one can still speak meaningfully of subsystems in relativistic quantum
theory. Although no new theorems will be proven in this paper, we
shall draw together results scattered through many highly technical
papers to make a coherent case for this claim. The technicalities will
be minimized as much as possible, however.

     What is a subsystem of a relativistic quantum system? We shall
not answer this question here. Indeed, as explained in Section
\ref{end}, even after the analysis carried out below, there are other
subtle matters to deal with before such a definition can be
attempted. But whatever a subsystem is, it is not merely a spatially
distinguished portion of the full system. To be conceptually most
useful, a subsystem should be an identifiable component of the system
which can subsist independently of the other subsystems comprising the
system, \eg it can be suitably screened off from the other subsystems
and studied experimentally without their influence. The analyzability
of the universe into subsystems therefore requires a concept of the
``independence'' of the subsystems, of which the relativistic quantum
world supports many distinct notions which coincide or are trivial in
the classical setting. The complex relation between these notions will
only be adumbrated here; the emphasis will be placed upon the warrant
for and the consequences of a particular notion of subsystem
independence, which, it is proposed, should be viewed as primary and
which, it is argued, provides a reasonable framework within which to
sensibly speak of relativistic quantum subsystems.

     In order to formulate in a mathematically rigorous manner the
notion of independent subsystems and to understand its consequences,
it is necessary to choose a mathematical framework which is
sufficiently general to subsume large classes of relativistic quantum
models, is powerful enough to facilitate the proof of nontrivial
assertions of physical interest, and yet is conceptually simple enough
to have a direct, if idealized, interpretation in terms of
operationally meaningful physical quantities. Such a framework is
provided by algebraic quantum field theory (AQFT) and algebraic
quantum statistical mechanics \cite{Ar,BrRo1,BrRo2,Haag}, also called
collectively local quantum physics, which is based on operator algebra theory,
itself initially developed by J. von Neumann for the express purpose
of providing quantum theory with a rigorous and flexible foundation
\cite{Neu,NeuI}.  This framework is briefly described in the next
section.

     In Section \ref{notions} we discuss three of the many notions of
independence which have been examined in the literature, indicating
briefly their operational meaning and their logical interrelations.
But what we regard as the operationally primary notion of independence
--- the split property --- is initially discussed in Section
\ref{split}.  This property is strictly stronger than all those
treated in Section \ref{notions}. After the somewhat abstract
discussion in Section \ref{split}, we present in Section
\ref{splitmeaning} a number of equivalent characterizations of the
split property which all have {\it operational} meaning. Further
physically significant consequences of the split property are reviewed
in Section \ref{further} to buttress our contention that the split
property should be viewed as the primary independence notion. Various
aspects of the warrant for the split property are considered in
Sections \ref{split}---\ref{further}.  Finally, in Section \ref{end}
we draw our conclusions and indicate why the analysis of the notion
of independent subsystems in relativistic quantum theory is far from
complete.

\section{Mathematical Framework}  \label{frame}

     The operationally fundamental objects in a laboratory are the
preparation apparata --- devices which prepare in a repeatable manner
the individual quantum systems which are to be examined --- and the
measuring apparata --- devices which are applied to the prepared
systems and which measure the ``value'' of some observable property of
the system. The physical notion of a ``state'' can be viewed as a
certain equivalence class of such preparation devices, and the
physical notion of an ``observable'' (or ``effect'') can be viewed as
a certain equivalence class of such measuring (or registration)
devices \cite{Lud,Ar}. In principle, therefore, these objects are
operationally determined, albeit quite abstract.

     It should be emphasized that these apparata can be effectively
extremely small, as exemplified by atomic traps or atomic
probes. Indeed, one probes the inner structure of protons and neutrons
by collisions with other suitable elementary particles. It requires no
undue stretch of the imagination to consider such collisions as being
part of the chain of either the preparation or the measuring apparata. 
Admittedly, such apparata are theory--dependent, but both
the design and the interpretation of {\it all} experiments are
strongly theory--dependent. We therefore see no immediate obstacle to
admitting the existence of apparata of submicroscopic extent. And,
although it is possible in principle to derive quantum theory without
any reference to microsystems by using only the description of
macrosystems in terms of suitable state spaces \cite{Lud0}, we shall
also posit the existence of microsystems and accept that quantum theory
describes (ensembles of) such systems.

     In algebraic quantum theory, such observables are represented by
self--adjoint elements of certain algebras of operators, either \W\ or
\C algebras.\footnote{For some purposes a more general formulation of
observable as a certain kind of positive operator valued measure is
useful \cite{Da,Ho,Lud,PeTe}. However, this more general class of
observables is subsumed into the present setting if the measure takes
its values in an algebra of the sort discussed here.} In this paper we
shall restrict our attention primarily to concretely represented \W
algebras, which are commonly called von Neumann algebras in honor of
the person who initiated their study \cite{NeuI}. The reader
unfamiliar with these notions may simply think of algebras $\Ms$ of
bounded operators on some (separable) Hilbert space $\Hs$ (or see
\cite{KaRi1,KaRi2,Tak1,Tak2,Tak3} for a thorough background).  We
shall denote by $\Bs(\Hs)$ the algebra of all bounded operators on
$\Hs$. Physical states are represented by mathematical {\it states} $\phi$,
\ie linear, (norm) continuous maps $\phi : \Ms \rightarrow \CC$ from
the algebra of observables to the field of complex numbers which take the
value 1 on the identity map $I$ on $\Hs$ and are positive in the sense
that $\phi(A^* A) \geq 0$ for all $A \in \Ms$. An important subclass
of states consists of {\it normal states}; these are states such that
$\phi(A) = \tr(\rho A)$, $A \in \Ms$, for some {\it density matrix}
$\rho$ acting on $\Hs$, \ie a bounded operator on $\Hs$ satisfying the
conditions $0 \leq \rho = \rho^*$ and $\tr(\rho) = 1$. A special case
of such normal states is constituted by the {\it vector states}: if
$\Phi \in \Hs$ is a unit vector and $P_\Phi \in \Bs(\Hs)$ is the
orthogonal projection onto the one dimensional subspace of $\Hs$
spanned by $\Phi$, the corresponding vector state is given by
$$\phi(A) = \langle \Phi, A \Phi \rangle = \tr(P_\Phi A) \, , \,
A \in \Ms \, .$$
Generally speaking, theoretical physicists tacitly restrict their
attention to normal states; for the purposes of this paper, it will
suffice to do so as well.

     In local quantum physics, one takes account of the localization
of the observables from the very outset. Indeed, in the context of
relativistic quantum field theory, since any measurement is carried
out in a finite spatial region and with a finite elapse of time, for
every observable $A$ there must exist bounded (open, nonempty)
spacetime regions $\Os$ in which $A$ can be localized.\footnote{It is
clear from operational considerations that one could not expect to
determine a minimal localization region for a given observable
experimentally. In \cite{Ku3} the possibility of determining such a
minimal localization region in the idealized context of AQFT is
discussed at length. In any case, from the remarks made above it is
clear that such localization regions can be very small indeed. And in
the idealized setting of quantum field theory, there are nontrivial
algebras of observables associated with every nonempty region $\Os$
--- see \eg \cite{DSW}.} We denote by $\As(\Os)$ the algebra generated by
all observables localized in $\Os$.  Clearly, it follows that if
$\Os_1 \subset \Os_2$, then $\As(\Os_1) \subset \As(\Os_2)$.  This
yields a net $\Os \mapsto \As(\Os)$ of observable algebras associated
with the experiment(s) in question. In turn, this net determines the
smallest algebra $\As$ containing all $\As(\Os)$. The preparation
procedures in the experiment(s) then determine states $\phi$ on $\As$,
the global observable algebra.
 
     The standard picture of a Hilbert space of state vectors familiar
from von Neumann's formulation of nonrelativistic quantum mechanics is then 
recovered as follows. A state $\phi$ on $\As$ uniquely determines 
(up to unitary equivalence \cite{KaRi1,Tak1}) the GNS representation 
$(\Hs_\phi,\pi_\phi,\Phi)$ of $\As$ with
$\pi_\phi : \As \rightarrow \Bs(\Hs_\phi)$ a representation of $\As$
as a concrete algebra of bounded operators acting on a Hilbert space
$\Hs_\phi$, $\Phi \in \Hs_\phi$ a cyclic vector for 
$\pi_\phi(\As)$,\footnote{\ie the set of 
vectors $\pi_\phi(A) \Phi$, $A \in \As$, is dense in $\Hs_\phi$} and,
for all $A \in \As$,
$$\phi(A) = \langle \Phi, \pi_\phi(A)\Phi\rangle \, .$$
The conceptual and mathematical advantages which accrue to the use of
this more general notion of state are manifold, but they will not play a
role in this paper.

     In the setting of relativistic quantum physics, it is argued that
spacelike separated events must be, in some sense, independent. Indeed, 
if two events are spacelike separated, there exists an inertial reference 
frame in which they are {\it simultaneous}. Since events which happen 
simultaneously cannot reasonably influence each other, each must be 
independent of the other (though they may have a common causal antecedent).
One way this independence is expressed in AQFT is through the property
of {\it locality} (sometimes referred to as microcausality or Einstein 
causality), namely if $\Os_1$ and $\Os_2$ are spacelike separated
regions, then $\As(\Os_1) \subset \As(\Os_2){}' = 
\{ B \in \Bs(\Hs) \mid AB - BA = 0 \, , \, 
\textrm{ for all} \quad A \in \As(\Os_2) \}$. However, as shall be explained 
below, in most concrete models much stronger forms of independence are 
satisfied by observable algebras associated with spacelike separated regions.  

     It should be remarked that the word ``locality'' is used in at
least two distinct ways in quantum theory. In nonrelativistic quantum
theory, the nonlocality spoken of when referring to Bell's
inequalities or entangled states in quantum information theory is a
property of certain {\it states} on the observable
algebras.\footnote{In the voluminous literature on Bell's inequality
there are yet other uses of the word ``local'' whose relations seem to
be uncharted.} In relativistic quantum field theory (and quantum
statistical mechanics), locality is a property of the {\it observable
algebras}. These two properties are perfectly compatible with each
other, as is evidenced by the fact that, quite generically, there
exist pairs of local algebras in relativistic quantum field models for
which Bell's inequalities are {\it maximally violated} in {\it all}
normal states \cite{SuWe5}, so that all such states are maximally
entangled (and therefore maximally ``nonlocal'') across such pairs,
even though the algebras themselves satisfy locality. 

\section{Some Formulations of Subsystem Independence}  \label{notions}

     There are various technical conditions used in algebraic quantum
theory to formulate the notion of the independence of subsystems. This
is only to be expected, since there are clearly different quantitative
and qualitative aspects of such independence. The study of these
formulations and their logical relations is therefore of some
conceptual interest.  In the context of classical mechanics or
classical field theory, these notions are either trivial or mutually
equivalent.  However, in the quantum setting they are distinct and
nontrivial. It is not our purpose here to review the multitude of
notions which have been under discussion in the literature nor to
explain their logical interrelationships (but see
\cite{Su2,FlSu,Red,Ham} for such reviews). Instead we shall restrict
our attention in this paper to just four of these. Two are commonly
employed in the theoretical physics literature, though only in the
context of a special case.

     Surely the most familiar formulation of the independence of
two subsystems with observable algebras $\As$ and $\Bs$, respectively,
is that their observables be mutually commensurable (or ``jointly
measurable'', ``mutually coexistent'', \etc.):

\medskip

\noindent {\bf Commensurability of Observables}: 
$[A,B] = AB - BA = 0$, for all $A \in \As$, $B \in \Bs$, \ie 
$\As \subset \Bs'$.

\medskip

\noindent  As this formulation is so familiar, we shall not elaborate upon
its operational significance, though such a discussion can be found in
\cite{Su2,Lud,BSS}. We remark that in the setting alluded to above, where
observables are modelled by positive operator valued measures, commutative
observables are jointly measurable (or mutually coexistent) but the 
converse is false \cite{La,LaPu,BSS} for certain pairs or finite families
of observables. Nonetheless, when concerned with independence of 
{\it subsystems}, one is clearly interested in the joint measurability
of {\it all} pairs of observables of the subsystems, and this {\it is}
equivalent to $\As \subset \Bs'$ (see \eg \cite{NeWe,Su2}).

     In \cite{HaKa} R. Haag and D. Kastler introduced a notion of
subsystem independence they named statistical independence. This is the 
property that each of the two subsystems can be prepared in any state,
independently of the preparation of the other system. In \cite{Su2}
this property was formalized for different classes of observable algebras
(and the distinctions proved to be apposite). Since we are restricting 
ourselves to von Neumann algebras in this paper and are not trying to
treat independence notions exhaustively, we shall only mention
the property termed \W independence in \cite{Su2}.

\medskip

\noindent {\bf Statistical Independence} (\W Independence): Let
$\As$ and $\Bs$ be von Neumann algebras acting on the Hilbert space
$\Hs$. The pair $\pair$ is statistically independent if for any normal 
state $\phi_1$ on $\As$ and any normal state $\phi_2$ on $\Bs$,
there exists a normal state $\phi$ on $\Bs(\Hs)$ such
that $\phi(A) = \phi_1(A)$ and $\phi(B) = \phi_2(B)$ for all
$A \in \As$, $B \in \Bs$.

\medskip

     One sees that if $\As$ and $\Bs$ represent the observable algebras
associated with two subsystems, the statistical independence of 
the pair $\pair$ can be loosely interpreted as follows: any two partial
states on the two subsystems can be realized by a suitable preparation
of the full system; no choice of a state prepared on one subsystem can prevent 
the other subsystem from being in any prescribed state.

     It turns out that commensurability of observables and statistical
independence are logically independent notions. Indeed, there exist
von Neumann algebras which do not mutually commute and yet are
statistically independent; and there exist mutually commuting von
Neumann algebras which are not statistically independent ---
cf. \cite{Su2} for a discussion, examples and further references.

     The third notion of subsystem independence we shall consider
here is of more recent origin \cite{ReSu3} and is closely related
to the fourth, and primary, notion to be discussed in the next
section. A few preparations shall prove to be necessary.

     We recall that a linear map $T \colon \As \to \Bs$ can be extended
to a linear map $T_n \colon M_n(\As) \to M_n(\Bs)$ (here $M_n(\As)$ is
the set of $n$ by $n$ matrices with elements from the algebra $\As$) by
$$T_n \left(%
\begin{array}{ccc}
  A_{11} & \ldots & A_{1n} \\
  . & . & . \\
  A_{n1} & \ldots & A_{nn} \\
\end{array}%
\right)=
\left(%
\begin{array}{ccc}
  T(A_{11}) & \ldots & T(A_{1n}) \\
  . & . & . \\
  T(A_{n1}) & \ldots & T(A_{nn}) \\
\end{array}%
\right) \, .
$$

     $T$ is said to be {\it completely positive} if $T_n$ is positive for 
every $n \in \IN$, \ie $T_n$ maps positive operators to positive operators. 
A completely positive map $T : \As \to \As$ satisfying 
$T(I) \leq I$ is called an {\em operation} \cite{Da,Kr}. An operation $T$ 
such that $T(I) = I$ is said to be {\it nonselective}. An operation $T$ on a 
von Neumann algebra $\As$ is called {\em normal\/} if it is 
$\sigma$--weakly continuous (the natural topology associated with von
Neumann algebras). A positive linear map $T : \As \to \Bs$
is {\it faithful} if $T(A) > 0$ whenever $\As \ni A > 0$.

     Operations are mathematical representations of physical
operations, \ie physical processes which take place as a result of
physical interactions with the quantum system. (For a more detailed discussion
of operations see \cite{Kr}.) A state on $\As$ is a completely
positive unit preserving map from $\As$ to $\CC$. So, if
$\phi$ is a state on $\As$, then 
\begin{equation}
\As \ni A \mapsto T(A) = \phi(A) I \in \As
\end{equation}
is a nonselective operation in the sense of the above
definition, which is canonically associated with the state and which
may be interpreted as the preparation of the system into the state
$\phi$. In fact, for any state $\omega$ and $A \in \As$ one has
$\omega(T(A)) = \omega(\phi(A)I) = \phi(A)$. Further examples of operations 
are provided by measurements. In particular, if one measures a quantum 
system with observable algebra
$\Bs(\Hs)$ for the value of a (possibly unbounded) observable $Q$ with
purely discrete spectrum $\{ \lambda_i \}$ and corresponding spectral
projections $P_i$, then according to the ``projection postulate" this
measurement can be represented by the operation $T$ defined as
\begin{equation} \label{measure}
\Bs(\Hs) \ni X \mapsto T(X) = \sum_iP_iXP_i \in \Bs(\Hs) \, .
\end{equation}
$T$ is a normal nonselective operation on $\Bs(\Hs)$. In fact, K. Kraus
proved that any normal operation $T$ on $\Bs(\Hs)$ must have the form
$$T(X) = \sum_i W_i^* X W_i  \qquad\qquad \sum_i W_i^* W_i \leq I \, , $$
where $W_i \in \Bs(\Hs)$ \cite{Kr}. It should also be mentioned that
in quantum information theory, unit preserving completely positive
linear maps $T : \As \rightarrow \Bs$ are called {\it channels} and
are of central interest \cite{Key}. 

     In the light of these considerations, the following generalization
of statistical independence is natural. Once again, we restrict our
attention to just one of the notions introduced in \cite{ReSu3}, in
this case to what was termed there operational \W independence in the 
product sense. Note that it is not assumed that the algebras mutually
commute.

\medskip

\noindent {\bf Operational Independence} (Operational \W Independence in the 
Product Sense): A pair $(\As,\Bs)$ of von Neumann
algebras is {\it operationally independent in $\As \vee \Bs$}, the smallest 
von Neumann algebra containing both $\As$ and $\Bs$, if any two (faithful) 
normal nonselective operations on $\As$ and $\Bs$, respectively, have a 
joint extension to a (faithful) normal nonselective operation on 
$\As \vee \Bs$ which factors across the pair; \ie if for 
any two (faithful) normal completely positive unit preserving linear maps
$$T_1 \colon \As \to \As \quad , \quad
T_2 \colon \Bs \to \Bs \, , $$
there exists a (faithful) normal completely positive unit preserving linear map
$$T \colon \As \vee \Bs \to \As \vee \Bs$$
such that
$$T(A) = T_1(A) \, , \, T(B) = T_2(B) \, , \, T(AB) = T(A) T(B) \, ,$$
for all $A \in \As$, $B \in \Bs$.

\medskip

     Hence, for operationally independent subsystems, {\it any}
operation performed on either subsystem is compatible with performing
{\it any} operation on the other subsystem in the strong sense that
{\it both} can be implemented by a single operation on the full system
which factors across the subalgebras. Operational independence of
$\pair$ is strictly stronger than statistical independence
\cite{ReSu3}, and we expect it to be logically independent of the
commensurability of observables. In the next section we consider a
yet stronger independence property.

\section{The Split Property}   \label{split}

     We turn now to the split property, an important
structure property of inclusions of von Neumann algebras, which has
been intensively studied for the purposes of both abstract operator
algebra theory and local quantum physics. We shall see that
it provides a particularly useful formalization of subsystem independence
and propose this as primary among notions of independence. In
the following $\As \overline{\otimes} \Bs$ denotes the (unique \W )
tensor product of two von Neumann algebras $\As$ and $\Bs$, which can
be thought of as acting upon $\Hs \otimes \Hs$ in the natural manner
$(A \otimes B)(\Phi \otimes \Psi) = A\Phi \otimes B\Psi$, for all
$A \in \As$, $B \in \Bs$, and $\Phi,\Psi \in \Hs$. A von Neumann algebra 
$\Ms$ is a factor if its center
$\Ms \cap \Ms'$ consists only of multiples of $I$. A factor is type I
if it is isomorphic to $\Bs(\Ks)$ for some Hilbert space $\Ks$. 
In general, a von Neumann algebra is type I if it is
isomorphic to the tensor product $\Bs(\Ks) \overline{\otimes} \Zs$,
where $\Zs$ is an abelian (\ie commutative) von Neumann algebra. Hence,
abelian von Neumann algebras are type I \cite{Tak1,KaRi2}.

\medskip

\noindent {\bf Split Property}: A pair $\pair$ of von Neumann algebras
is {\it split} if there exists a type I factor $\Ms$ such that
$\As \subset \Ms \subset \Bs{}'$.

\medskip

\noindent Although according to the usage introduced in \cite{DoLo} we
should say that the pair $(\As,\Bs')$ is split, it is for our purposes
more convenient to use the terminology established above.  It is
immediately clear that mutually commuting type I factors are split.
The split property is equivalent to a structure property which may be
more familiar to the reader.

\begin{theorem} [\cite{Bu,DaLo}] \label{detlev}
For a mutually commuting pair 
$\pair$ of von Neumann algebras acting on a Hilbert space $\Hs$, the 
following are equivalent.

   1. The pair $\pair$ is split.

   2. The map
$$AB \to A \otimes B \, , \, A \in \As \, , \, B \in \Bs $$
extends to a spatial isomorphism of $\As \vee \Bs$ with 
$\As \overline{\otimes} \Bs$, \ie there exists a unitary 
operator $U : \Hs \to \Hs \otimes \Hs$ such that 
$U AB U^* = A \otimes B$ for all $A \in \As$, $B \in \Bs$.

\end{theorem}

     If $\pair$ is split, then it is operationally independent, thus also
statistically independent, and $\As \subset \Bs{}'$, but the
converse is false, \ie the split property is strictly stronger than
any combination of the other three \cite{Su2,DaLo,ReSu3}. 

     Before we explore various operational consequences of the split
property, let us examine the status of these notions of independence in 
relativistic quantum field theory. First of all, being spatiotemporally
distinct does {\it not} entail any kind of independence of the corresponding
observable algebras at all. In particular, even if 
$\Os_1 \cap \Os_2 = \emptyset$, all four independence 
properties can be false. This is not surprising in light of the causal
propagation of influences in relativistic quantum field theory ---
cf. \eg \cite{ReSu1}. 

     As mentioned above, if $\Os_1$ and $\Os_2$ are spacelike separated
(this entails $\Os_1 \cap \Os_2 = \emptyset$), then 
$\As(\Os_1) \subset \As(\Os_2){}'$ is satisfied in essentially all
constructed quantum field models. Indeed, the property of locality
is such a {\it sine qua non}, that it is normally posited as an axiom
of AQFT. In addition, in typical models the pair $\pairf$ is statistically
independent whenever $\Os_1$ and $\Os_2$ are spacelike separated 
\cite{FlSu}. However, when such spacelike separated regions are
tangent, \ie their closures have nonempty intersection, then
$\pairf$ is neither operationally independent nor split \cite{SuWe5,ReSu3}.

     But when $\Os_1$ and $\Os_2$ are {\it strictly} spacelike
separated, \ie when $\Os_1 + \Ns$ is spacelike separated from $\Os_2$
for some neighborhood $\Ns$ of the origin in $\RR^4$, then it is
typically the case that the pair $\pairf$ is split. Indeed, the split
property has been verified for all pairs associated with strictly
spacelike separated (precompact, convex) regions $\Os_1,\Os_2$ in a
number of physically relevant quantum field models, both interacting
and noninteracting \cite{Bu,Su1,BSads,DaLo,Le2}. There do exist models
in which the spacelike separation between $\Os_1$ and $\Os_2$ must
exceed a certain minimum bound before the corresponding pair $\pairf$
is split ({\it distal split property}) \cite{DDFL}, but the only known
models in which even the distal split property does not hold are
physically pathological models, such as models with noncompact global
gauge group and models of free particles such that the number of
species of particles grows rapidly with mass \cite{DoLo}. It should be
remarked that in more than two spacetime dimensions, pairs $\pairf$
associated with certain spacelike separated regions which are
unbounded (such as wedge regions) cannot be split, no matter how large
the spacelike separation between them may be \cite{Bu}. Nonetheless,
under fairly general circumstances observable algebras associated with
certain smaller but still unbounded regions called spacelike cones are
split when the regions are strictly spacelike separated \cite{DaFr}.

     Prospective readers of the above--mentioned papers should note
that the ``split property'' which is proven there is actually stronger
than the ``split property'' we defined above. Indeed, in the first decades
of the development of AQFT the property which is verified in the cited
papers was called the {\it funnel property}.  The terminology seems to
be in flux now. In any case, we explain the connection between the
funnel property and the property we term the split property. To
minimize technical complications we consider only spacetime regions
called double cones. These are nonempty intersections of some forward
light cone with some backward light cone \cite{Ar,Haag}. They are
convex, precompact and satisfy $(\Os{}')' = \Os$. The funnel property
obtains when for any two double cones $\Os$, $\widetilde{\Os}$ such
that the closure of $\Os$ is contained in $\widetilde{\Os}$, there
exists a type I factor $\Ms$ such that 
$\As(\Os) \subset \Ms \subset \As(\widetilde{\Os})$. Thus, the pair
$(\As(\Os),\As(\widetilde{\Os}){}')$ is split in our sense. If
$\Os_1$ is strictly spacelike separated from $\Os_2$, then there exists
a double cone $\widetilde{\Os}$ containing the closure of $\Os_1$
such that $\Os_2 \subset \widetilde{\Os}{}'$. By locality, one has 
$\As(\Os_2) \subset \As(\widetilde{\Os})'$. The funnel property therefore 
entails that the pair $(\As(\Os_1),\As(\Os_2))$ is split in our sense. 
The apparent loss of the distinguishing term funnel property is regrettable. 
Similarly, in the above papers the distal split property refers to the 
requirement that there is a minimal distance between the boundaries of $\Os$ 
and $\widetilde{\Os}$ before there exists a type I factor $\Ms$ such that
$\As(\Os) \subset \Ms \subset \As(\widetilde{\Os})$.

     In addition to {\it this} evidence that the split property is
commonly satisfied by physically relevant quantum field models, there
is further support for this hypothesis. The split property for all
pairs $\pairf$ associated with strictly spacelike separated
(precompact, convex) regions $\Os_1,\Os_2$ (in fact, the funnel
property) has been shown \cite{BuWi,BDF,BDL2} to be a consequence of a
condition (nuclearity) which assures that the model is
thermodynamically well--behaved (\eg thermal equilibrium states exist
for all temperatures \cite{BuJu1,BuJu2}).  The nuclearity condition,
which necessitates rather heavy technical baggage and will therefore
not be explained here, expresses the requirement that the
energy--level density for any states essentially localized in a
bounded spacetime region cannot grow too fast with the energy.  There
is good physical reason to expect that physically relevant models will
satisfy this condition and therefore also the split property
\cite{BuWi,BDL2}.

     Although both the split property and the good thermodynamic
behavior are consequences of the nuclearity condition (and not
necessarily the converse\footnote{However, in \cite{BDL1} the split
property is proven to be equivalent to the nuclearity of a certain
family of maps and certain other partial converses are established in
\cite{BDL2}.}), the existence of a minimal spacelike separation for
which the split property holds (the distal split property) has been
shown to be related to the existence of a maximal temperature above
which thermal equilibrium states do not exist \cite{BuYng,BDL2}. In
addition, all known models which do not satisfy at least the distal
split property have in common that they describe systems with an
enormous number of local degrees of freedom.  These systems, as a
consequence, do not admit thermal equilibrium states, \eg
\cite{BuJu1}.  Hence, there are at least some indications that the
split property is directly related in some not yet fully understood manner to
good thermodynamic properties of quantum fields.

     We have already pointed out the fact that commuting type I
factors are always split; indeed, for type I factors 
$\As \subset \Bs{}'$ is {\it equivalent} to $\pair$ is split. In 
nonrelativistic quantum mechanics the
observable algebras are typically type I. Moreover, in the now
extensive literature on quantum information theory (developed up to the
present primarily for nonrelativistic quantum theory and quantum systems of
only finitely many degrees of freedom) the word
``subsystem'' has become synonymous with a type I observable algebra which
is a factor in a tensor product of other ``subsystems'' --- see \eg
\cite{Key}. Hence, the split property is the second of the
above--mentioned two independence properties commonly used by
theoretical physicists. However, it is employed by them in a setting where 
the most distinctive advantages of the property cannot reveal themselves.

\section{Physical Characterizations of the Split Property} 
\label{splitmeaning}

     In light of the above, one may tentatively conclude that the
split property obtains in some generality in physically relevant
quantum field models. We further examine the warrant for this property
by explaining some physically meaningful characterizations of the
split property.  We begin with one of the first found, which
generalized a characterization proven in \cite{BDL}. We present it 
in a form given in \cite{Su2}, since the
original \cite{We} requires the full apparatus of AQFT, the
minimal assumptions of which entail that local algebras are ``almost''
type III \cite{Bor}. (In fact, with some additional hypotheses ---
which nonetheless are also commonly fulfilled by most concrete models
--- local observable algebras {\it are} type III, see \eg \cite{Fred}.) 
A characterizing property of type III factors (and the condition
actually needed in the proof of the following theorem) is that in such an
algebra $\Ms$ for any nonzero projection $P \in \Ms$ there exists a 
partial isometry $W \in \Ms$ such that $WW^* = P$ and $W^* W = I$.
Thus all nonzero projections in type III algebras are infinite-dimensional 
\cite{Tak1, KaRi2}.\footnote{The readers of this 
journal might find the introduction to types of von Neumann algebras given 
in \cite{ReSu2} relatively painless.}

\begin{theorem} [\cite{We,Su2}] \label{reinhard} Let $\As$,
$\Bs$, be commuting type III von Neumann factors on the Hilbert space
$\Hs$. Then the following are equivalent:

   1. The pair $\pair$ is split.

   2. Local preparability of some normal state: there exists a normal
state $\phi$ and a normal positive map $T : \Bs(\Hs) \rightarrow \Bs(\Hs)$
such that $T(A) = \phi(A) T(1)$ for all $A \in \As$ and $T(B) = T(1)B$
for all $B \in \Bs$.

   3. Nonselective local preparability of all normal states: for any normal
state $\phi$ there exists a map $T : \Bs(\Hs) \rightarrow \Bs(\Hs)$ of
the form $T(C) = \sum W_i^* C W_i$ with $W_i \in \Bs'$ such that 
$T(A) = \phi(A) T(1)$ for all $A \in \As$ and $T(1) = 1$.

\end{theorem}

     So for any state $\omega$ and all $A \in \As$, $B \in \Bs$, 
\begin{eqnarray*}
\omega(T(AB)) & = & \omega(\sum W_i^* AB W_i) = \omega(B \sum W_i^* A W_i)
= \omega(B T(A)) \\
& = & \phi(A)\omega(B) \, .
\end{eqnarray*}
$\omega \circ T$ is therefore a product state on $\As \vee \Bs$. Thus
normal product states with arbitrary normal partial states can be
locally prepared on $\As \vee \Bs$ if and only if $\pair$ is split,
whenever the algebras are type III.

     In view of the previous discussion of operations, the operational
content of conditions (2) and (3) in Theorem \ref{reinhard} should be
clear. Note that the operations involved in the above theorem are
solely operations to prepare states. The following result has been
recently proven and involves, as discussed in Section \ref{notions},
arbitrary operations.

\begin{theorem} [\cite{ReSu3}] \label{new} 
If the pair $\pair$ is split, then it is operationally independent. Moreover, 
if either of the algebras is type III or a factor, then the pair $\pair$ is 
operationally independent if and only if it is split.
\end{theorem}

     In fact, it is proven in \cite{ReSu3} that the version of operational
independence discussed here is equivalent to a property called
\W independence in the product sense. And, as shown in \cite{DaLo},
in a large number of circumstances (though not all) \W independence
in the product sense is equivalent to the split property. The two 
particular circumstances mentioned in Theorem \ref{new} are singled
out, because they arise frequently in applications to quantum theory. 

     We turn next to another, relatively recently discovered
characterization of the split property which has operational
interpretation.  A bit of background information might be useful
here. As mentioned above, the commensurability of observables and
statistical independence are independent properties. Although some
rather {\it ad hoc} and nontransparent conditions expressed in terms
of states on algebras $\As$, $\Bs$, are known which entail that 
$\As \subset \Bs{}'$ (see \cite{Su2} for a discussion of these as well as
references), workers in the field have long desired operationally
meaningful conditions on the states on $\As$, $\Bs$, which would imply
that the algebras mutually commute. This search was taken up in
\cite{BS}, but what was found was a characterization of the split property.

     Let $\As$, $\Bs$, be two observable algebras which need not commute,
and let $E \in \As$, $F \in \Bs$, be projections. Such projections
represent ``yes--no'' observables, such as ``the spin of the electron
in the $z$-direction is up--down'' or ``the polarization of the photon
is right--handed---left--handed''. What kind of meaningful coincidence
experiment can be designed in the case that $EF \neq FE$? Let $E \wedge F$
denote the largest projection in $\Bs(\Hs)$ dominated by both $E$ and
$F$. Note that $E \wedge F = EF$ if and only if $EF = FE$. Moreover,
$$E \wedge F = \lim_{n \rightarrow \infty} (EF)^n = 
\lim_{n \rightarrow \infty} (FE)^n \, .$$
In a suitable coincidence experiment involving the observables $E$ and $F$, 
a ``yes'' result in the apparatus should yield ``yes'' with certainty
for any subsequent measurement of either $E$ or $F$; moreover, the
acceptance rate of the device should be maximized by the design of the
experiment. A bit of thought convinces one that $E \wedge F$ is the
(idealized) observable corresponding to this optimized coincidence apparatus.

\begin{definition}
A state $\omega$ on $\As \vee \Bs$ is $\As$-$\Bs$--uncorrelated if
$\omega(E \wedge F) = \omega(E) \omega(F)$ for all projections
$E \in \As$ and $F \in \Bs$.
\end{definition}

     Note that if $\As \subset \Bs{}'$, then in such an 
$\As$-$\Bs$--uncorrelated state one would have 
$\omega(EF) = \omega(E) \omega(F)$ for all projections
$E \in \As$ and $F \in \Bs$. This would in turn entail that
$\omega(AB) = \omega(A) \omega(B)$ for all $A \in \As$ and $B \in \Bs$
(by the spectral theorem and the fact that $\As$ and $\Bs$ are \C algebras).
But this latter condition corresponds exactly to the notion of independence 
widely used in classical probability theory (and is
a direct generalization of it, cf. \eg \cite{ReSu2}). However, the
condition is also operationally meaningful when the algebras do not
commute. Indeed, this fact allows one to define a quantitative
measure of the degree of noncommutativity of two such algebras \cite{BS}.
Instead, we return to our immediate purpose.

\begin{theorem} [\cite{BS}] Let $\As$, $\Bs$, 
be two von Neumann algebras which need not commute (and $\As \vee \Bs$ be
a factor). Then $(\As,\Bs)$ is split if and only if
there exists a normal $\As$-$\Bs$--uncorrelated state on $\As \vee \Bs$.
\end{theorem}

     The existence of a single normal $\As$-$\Bs$--uncorrelated state on 
$\As \vee \Bs$ suffices to entail that $(\As,\Bs)$ is split. However,
if $(\As,\Bs)$ is split, then $\As \subset \Bs{}'$, so that 
$E \wedge F = EF$ for all projections $E \in \As$, $F \in \Bs$. In addition,
for any normal states $\phi_1$ on $\As$ and $\phi_2$ on $\Bs$, there exists
a normal state $\phi_1 \otimes \phi_2$ on $\As \overline{\otimes} \Bs$
such that 
$$ (\phi_1 \otimes \phi_2)(A \otimes B) = \phi_1(A) \phi_2(B) \, , $$
for all $A \in \As$ and $B \in \Bs$. Defining 
$\psi : \As \vee \Bs \rightarrow \CC$ by
$$\psi(AB) =  (\phi_1 \otimes \phi_2) (UABU^*) $$
(and extending to $\As \vee \Bs$ by linearity and continuity), where $U$ 
is the unitary from Theorem \ref{detlev}, one sees that
\begin{eqnarray*}
\psi(E \wedge F) & = & \psi(EF) = (\phi_1 \otimes \phi_2) (UEFU^*) \\
& = & (\phi_1 \otimes \phi_2) (E \otimes F) = \phi_1(E)\phi_2(F) \\
& = & \psi(E)\psi(F) \, ,
\end{eqnarray*}
for all projections $E \in \As$, $F \in \Bs$. In other words, if
$\pair$ is split then any normal partial states on the subalgebras
have extensions to normal $\As$-$\Bs$--uncorrelated states on $\As \vee \Bs$.

     The three theorems in this section provide different operational
characterizations of the split property. In the following section we
shall discuss some physically relevant consequences of the split
property with an eye towards those which shed further light upon its
relevance to subsystems.

\section{Further Consequences of the Split Property}  \label{further}

     In general, in relativistic quantum field theory one has global energy, 
momentum and charge observables (say $Q$) which have meaning for the full
quantum system \cite{Ar,Haag}. These cannot be localized in any region with 
nonempty causal complement and cannot directly refer to any subsystem. But 
to any subsystem worth the name one must be able to attribute such quantities. 
This is a highly nontrivial matter, but if the funnel property 
holds\footnote{As mentioned above, also strictly spacelike separated spacelike 
cones are associated with split observable algebras in some generality. For 
this reason, the results discussed here are also valid for charges which can 
only be localized in such spacelike cones and not in bounded regions, as is 
expected in massive gauge theories \cite{DaFr}.}, then for any double cone 
$\Os$ and any slightly larger double cone $\widetilde{\Os}$ there exist 
observables in $\As(\widetilde{\Os})$ (say 
$Q_{\widetilde{\Os}}$)\footnote{If $Q$ is an unbounded selfadjoint operator, 
then so is $Q_{\widetilde{\Os}}$, which entails 
$Q_{\widetilde{\Os}} \notin \As(\widetilde{\Os})$. Nonetheless, 
$\As(\widetilde{\Os})$ does contain all of the spectral projections of 
$Q_{\widetilde{\Os}}$.} which are indistinguishable from the corresponding 
global observables for any experiment implementable in $\Os$:
\begin{equation} \label{charge}
e^{itQ} A e^{-itQ} = e^{itQ_{\widetilde{\Os}}} A e^{-itQ_{\widetilde{\Os}}}
\, , 
\end{equation}
for all $A \in \As(\Os)$ and all $t \in \RR$, if $Q$ is a generator of
a global gauge group, or for all $t$ in a suitable neighborhood of $0$
if $Q$ is a global energy or momentum operator (see below)
\cite{BDL,Do,DoLo0,DoLo}.\footnote{For the reader who knows
that the action of gauge groups upon the observables is trivial, we note that 
the same relation (\ref{charge}) holds when the net of observable algebras 
$\Os \mapsto \As(\Os)$ is replaced by a net of field algebras
$\Os \mapsto \Fs(\Os)$ \cite{BDL}, upon which the action {\it is} nontrivial.}
In fact, if the global theory is supplied with a strongly continuous
unitary representation $U(\Pid)$ of the identity component of the
Poincar\'e group acting covariantly upon the observable algebras
$$U(\lambda) \As(\Os) U(\lambda)^{-1} = \As(\lambda\Os) \, , $$
for all $\lambda \in \Pid$ and $\Os$, then for a fixed spacetime region
$\Os$, any neighborhood $\Ps_0$ of the identity in $\Pid$ and any
region $\widetilde{\Os}$ such that $\lambda \Os \subset \widetilde{\Os}$
for all $\lambda \in \Ps_0$, one has unitaries 
$U_{\widetilde{\Os}}(\lambda) \in \As(\widetilde{\Os})$ such that 
$$U_{\widetilde{\Os}}(\lambda) A U_{\widetilde{\Os}}(\lambda)^{-1} =
U(\lambda) A U(\lambda)^{-1} \, ,$$
for all $\lambda \in \Ps_0$ and $A \in \As(\Os)$.
In addition, the local implementers $Q_{\widetilde{\Os}}$ have the {\it same}
spectrum as their global counterparts \cite{BDL}. For example, if
the generators of the translation subgroup $U(\RR^4) \subset U(\Pid)$
(which have the interpretation of the global energy--momentum observables of
the quantum field theory) satisfy the relativistic spectrum condition, then 
so do the generators of the local implementers $U_{\widetilde{\Os}}(x)$, 
$x \in \RR^4$. Hence, subsystems whose observables are localized 
in $\Os$ can be attributed localized energy--momentum and charge operators, 
at the minor cost of accepting a slightly larger localization region for the 
latter.\footnote{In light of the observation made above that there is no
operational way to determine the minimum spacetime localization of an 
observable, this price is minor, indeed.} The arguments yielding these results 
can also be applied to supersymmetric theories and theories in which 
topological charges are present \cite{BDL}. 

     We turn finally to some significant consequences of the split property
which are not immediately relevant to subsystem independence, but which
will be very briefly mentioned to further illustrate to the reader the 
mathematical and conceptual advantages of the split property. We do not 
attempt to provide an exhaustive list of the known or suspected 
consequences of this kind. 

     In general, in relativistic quantum field theory quantities such
as entropy, entanglement of formation and relative entropy of
entanglement are infinite (undefinable).  If the funnel property (or
nuclearity, which entails the funnel property) holds
for strictly spacelike separated regions, then such quantities can be
given sensible (strictly positive and finite) meaning at least for a
dense subset of normal states \cite{Na1,Na2}.

     As mentioned above, the local algebras in AQFT are typically type III
(in fact, type III$_1$ in the classification of A. Connes \cite{Fred}).
However, there exist nonisomorphic type III$_1$ algebras. A further 
consequence of the funnel property is that essentially {\it all} local 
observable algebras are isomorphic \cite{BDF}. This makes it unmistakeably 
obvious that it is not the algebras of observables themselves in which are
encoded most of the physical information of the theory but rather the
{\it inclusions} of the local algebras, \ie how the algebra $\As(\Os_1)$
sits inside the algebra $\As(\Os_2)$ when $\Os_1 \subset \Os_2$. 

     It has also emerged from recent work \cite{BuLe,BSads,Le2,Le3} that
the funnel property is exceedingly useful in the rigorous construction
of quantum field models using algebraic methods. For a relatively
nontechnical overview of the results obtained with these methods, as
well as further references, see \cite{construction}.

     The split property can also be employed to resolve other
conceptual problems in quantum field theory. An example is the fairly
recent exchange in which G.C. Hegerfeldt argued that there are
causality problems in Fermi's classic two--atom system \cite{He}. But
D. Buchholz and J. Yngvason convincingly retorted that Hegerfeldt's
argument rested upon his tacit use of a local, minimal
projection.\footnote{A minimal projection of a von Neumann algebra
$\Ms$ is a projection in $\Ms$ which dominates no other projection in
$\Ms$ other than the zero operator.}  But local observable algebras
$\As(\Os)$ are type III algebras, which contain no minimal
projections. (Indeed, as mentioned above, in type III algebras all
nonzero projections are mutually equivalent in a sense which places them as
far away from minimal projections as is possible in the theory of
operator algebras.)  On the other hand, the type I algebras which
interpolate between local algebras in the split property {\it do} have
minimal projections.  Buchholz and Yngvason explain how once this
point is taken into account, the causality problems Hegerfeldt
described evaporate \cite{BuYng2}.

\section{Concluding Remarks}  \label{end}

    We conclude that it is meaningful to speak of independent subsystems in
relativistic quantum theory, {\it if} they can be localized in
spacetime regions $\Os_1$, resp. $\Os_2$, such that $\As(\Os_1)$ and
$\As(\Os_2)$ satisfy the split property. For then their observables
are mutually commensurable, they can be independently and locally
prepared in arbitrary states, they are ``operationally independent,''
and they possess mutually compatible localized energy, momentum and
charge observables, to mention just a few desirable properties.  In addition,
the split property for spacetime regions which are (sufficiently)
strictly spacelike separated holds widely in the quantum field models
constructed up to this point and is expected to hold in most, if not all,
physically relevant models to be constructed in the
future. Syllogistically, it is therefore meaningful to speak of
independent subsystems in relativistic microscopic physics.

     These observations are, however, only a starting point for the
examination of the notion of such subsystems. No attempt has been made
here to provide a definition of microscopic subsystem, although it is
clear that this is an interesting philosophical question. We briefly
indicate some of the complexities involved in coming to grips with
this question in relativistic quantum physics. 

     To begin, let us consider the notion of quarks, which are
essential to the understanding of the physics of strong interactions
in the Standard Model of elementary particle physics. Although there
are indirect indications that something like quarks exist within
baryons such as protons and neutrons, can one correctly speak of them
as subsystems? According to heuristic computations made in the
Standard Model, quarks can never be observed as isolated objects,
since the force of attraction between coupled quarks grows with their
separation. This is antithetical to the usual situation one faces when
analyzing nature into subsystems, where the interparticle forces
decrease as the systems are mutually separated. The quarks which are
imagined to be bound together in colliding baryons can exchange
partners but, apparently, cannot be isolated as independent
subsystems, since each quark is always in intense interaction with at
least one other quark. Are such objects subsystems? Are they more than
convenient theoretical constructs? Do subsystems need to be ``real''?

     These questions seem to us to be nontrivial. We remark that
quarks can possibly be understood in an intrinsic, objective manner
as ultraparticles \cite{BPS,Bu2,Bu3}, but it is not clear that this
(theoretically significant) observation diminishes the burden of
the above--mentioned philosophical questions.

     To indicate further the nontrivial nature of the question at
hand, we turn to the notion of ``particle'' in relativistic quantum
field theory, a subject receiving increasing attention from
philosophers of physics. Though some of these have chosen to deny the
existence of such particles (\eg \cite{Fr}) and others have argued,
with reservations, in their favor (\eg \cite{HaCl}), no one can deny
their centrality in the discourse and {\it Weltanschauung} of quantum
field theorists.  Certainly, it is already evident that the
comfortable notion of particles familiar from classical theory is
quite out of place in relativistic quantum physics; this is true even
before more subtle notions of particles which are in use among
mathematical physicists, such as infraparticles and ultraparticles,
have even entered the discussion among philosophers of physics.

     But {\it something} is being observed at large times and
distances (relative to collision times and distances) in scattering
experiments in CERN and elsewhere which acts much like particles
should, and whether one views these as particles, local excitations of
the quantum field or some other alternative, one must ask if these can
and should be viewed as (independent) subsystems. Certainly,
elementary particle physicists are acting as if they were. But it has
been rigorously established in AQFT that, whatever these things are,
they (or, more precisely, the idealized apparata which count them) are
not strictly localized, \ie cannot be elements of a local algebra
$\As(\Os)$ with $\Os{}' \neq \emptyset$ \cite{Haag,Ar}. All is not lost,
however, since they can be arbitrarily well approximated in norm by 
strictly localized observables \cite{Haag,Ar,Bu2} (see also \cite{HaCl}
for an explication written for philosophers of physics), but this is an
additional complication in the problem of deciding if these 
particle--like objects can be viewed as independent subsystems, even
asymptotically at large times and distances.

\end{document}